\documentclass[12pt]{article}
\usepackage{latexsym,amsmath,amssymb,amsbsy,graphicx}
\usepackage[cp1251]{inputenc}
\usepackage[T2A]{fontenc}
\usepackage[russian,english]{babel}
\usepackage[space]{cite} 

\begin{document}
\bigskip

{\bf

\centerline{{Wavelet-analysis of series of observations of relative
sunspot numbers.}} \centerline{{The dependence of the periods of
cyclic activity on the time }}
 \centerline{{at different time scales }}}

\bigskip

\centerline {A.A.~Borisov$^{1}$, E.A.~Bruevich$^{2}$, V.V.~
Bruevich$^{3}$,}
\centerline {I.K.~Rozgacheva$^{4}$, E.V.
Shimanovskaya$^{5}$}

\centerline {\it $^{2, 3, 5}$
Sternberg~Astronomical~Institute,~Moscow~ State~ University,}
\centerline {\it Universitetsky~pr., 13,~Moscow 119992,~Russia}

\centerline {\it $^{1, 4}$ VINITY ~RAS,~Moscow,~Russia}\

\centerline {\it e-mail:}
{$^{1}$borisov.andrey03@gmail.com,$^{2}$red-field@yandex.ru,$^{3}$brouev@sai.msu.ru,}

\centerline {$^{4}$rozgacheva@yandex.ru, $^{5}$eshim@sai.msu.ru} \

\bigskip

{\bf Abstract.} We applied the method of continuous
wavelet-transform to high-quality time-frequency analysis to the
sets of observations of relative sunspot numbers. Wavelet analysis
of these data reveals the following pattern: at the same time there
are several activity cycles whose periods vary widely from the quasi
biennial up to the centennial period. These relatively low-frequency
periodic variations of the solar activity gradually change the
values of  periods of different cycles in time. This phenomenon can
be observed in every cycle of activity.

\bigskip

\vskip12pt
\section{Introduction}
\vskip12pt

It's
known that the various activity indices which characterized the
different aspects of the solar magnetic activity correlate quite
well with the most popular solar index such as the relative sunspot numbers (SSN)
and with others indices over long time scales.

The relative sunspot numbers is a very popular, widely used solar
activity index: the series of relative sunspot numbers direct
observations continue almost two hundred years.

First we have studied yearly averaged values of SSN during  activity
cycles 1 - 23, the tree-hundred years data set we demonstrate on
Figure 1. We use the data from NGDC web site available at
http://www.ngdc.noaa.gov/stp and combined observational data from
National Geophysical Data Center Solar and Terrestrial Physics,
Solar-Geophysical Data Bulletin (2012) and Reports of National
Geophysical Data Center Solar and Terrestrial Physics (2012).

 On Figure 2 we illustrated with help of wavelet - analysis the fact
that the long time series of observations give us the very useful
information for study of the problem of solar flux cyclicity on long
time scales. The result of wavelet - analysis of series of
observations of average annual SSN is presented in form of many
isolines. For the isoline of the value of the wavelet-coefficients
are of the same. The maximum values of isolines specify the maximum
values of wavelet-coefficients, which corresponds to the most likely
value of the period of the cycle.  On Figure 2 there are three
well-defined cycles of activity: - the main cycle of activity is
approximately equal to a 10 - 11 years, 40-50- year cyclicity and
100 to 120-year (ancient) cyclicity.

Then we have studied monthly averaged values of SSN during  activity
cycles 21, 22 and 23. We also use the  data set from NGDC web site
available at http://www.ngdc.noaa.gov/stp.

Floyd $\it{et~al.}$ (2005) showed that the mutual relation between
sunspot numbers and three solar UV/EUV indices, and also with the
$F_{10,7}$ flux and the Mg II core-to-wing ratio remained stable for
25 years of satellite EUV-observations. Other solar activity indices
are also closely correlated to the SSN and radio flux $F_{10,7}$,
see Bruevich $\it{et~al.}$ (2014a). Some physics-based models have
been developed with using the combined proxies describing sunspot
darkening (sunspot number or areas) and facular brightening (facular
areas, CaII or MgII indices), see Viereck et al. (2001), Krivova et
al. (2003), Viereck et al. (2004) , Skupin et al. (2005).

The SSN index has an advantage over other indices of activity
because data on annual variation available from the 1700's.

The magnetic activity of the Sun is called the complex of
electromagnetic and hydrodynamic processes in the solar atmosphere
and in the underphotospheric convective zone, see Rozgacheva and
Bruevich (2002), Bruevich et al. (2001). The analysis of active
regions (plages and spots in the photosphere, flocculae in the
chromosphere and prominences in the corona) requires to study the
magnetic field of the Sun and the physics of magnetic activity. This
task is of fundamental importance for astrophysics of the Sun and
the stars. Its applied meaning is connected with the influence of
solar active processes on the Earth's magnetic field.

\vskip12pt
\section{Wavelet-analysis of series of observations of SSN}
\vskip12pt

The study indices of solar activity are very important not only for
analysis of solar radiation which comes from different altitudes of
solar atmosphere. The most important for solar-terrestrial physics
is the study of solar radiation influence on the different layers of
terrestrial atmosphere (mainly the solar radiation in EUV/UV
spectral range which effectively heats the thermosphere of the Earth
and so affects to our climate).

The relative sunspot number is an index of the activity of the
entire visible disk of the Sun. The SSN is a commonly used index of
solar activity, Vitinsky et al. (1986). Sunspots are temporary
phenomena on the photosphere of the Sun that appear visibly as dark
spots compared to surrounding regions. They are caused by intense
magnetic activity, which inhibits convection by an effect comparable
to the eddy current brake, forming areas of reduced surface
temperature. Although they are at temperatures of roughly 3000-4500K
(2700 - 4200°C), the contrast with the surrounding material at about
5,780 K (5,500°C) leaves them clearly visible as dark spots.
Manifesting intense magnetic activity, sunspots host secondary
phenomena such as coronal loops(prominences) and reconnection
events. Most solar flares and coronal mass ejections originate in
magnetically active regions around visible sunspot groupings.
Similar phenomena indirectly observed on stars are commonly called
stars pots and both light and dark spots have been measured.

We have to point out that close interconnection between radiation
fluxes characterized the energy release from different atmosphere's
layers is the widespread phenomenon among the stars of late-type
spectral classes, see Bruevich (2015a). It was confirmed that there
exists the close interconnection between photospheric and coronal
fluxes variations for sun-like stars of F, G, K and M spectral
classes with widely varying activity of their atmospheres, see
Bruevich and Alekseev (2007), Bruevich et al. (2001). It was also
shown that the summary areas of spots and values of X-ray fluxes
increase gradually from the sun and sun-like HK project stars with
the low spotted discs to the highly spotted K and M-stars. The main
characteristic describing the photospheric radiation is the
spottiness of the stars. Thus, the study of the relative sunspot
numbers is very important to explain the observations of sun-like
stars.

\begin{figure}[tbh!]
\centerline{
\includegraphics[width=140mm]{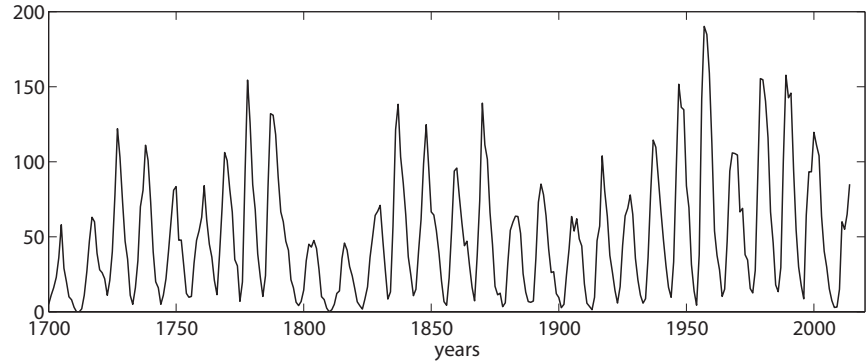}}
 \caption{Yearly averaged relative sunspot numbers 1700 - 2015.}
{\label{Fi:Fig1}}
\end{figure}

In Fig.1 we can see that the duration of the 11 yr cycle of solar
activity ranged from 7 to 17 years. The results become more accurate
with the beginning of the of direct solar observations (1850-2015).

\begin{figure}[tbh!]
\centerline{
\includegraphics[width=140mm]{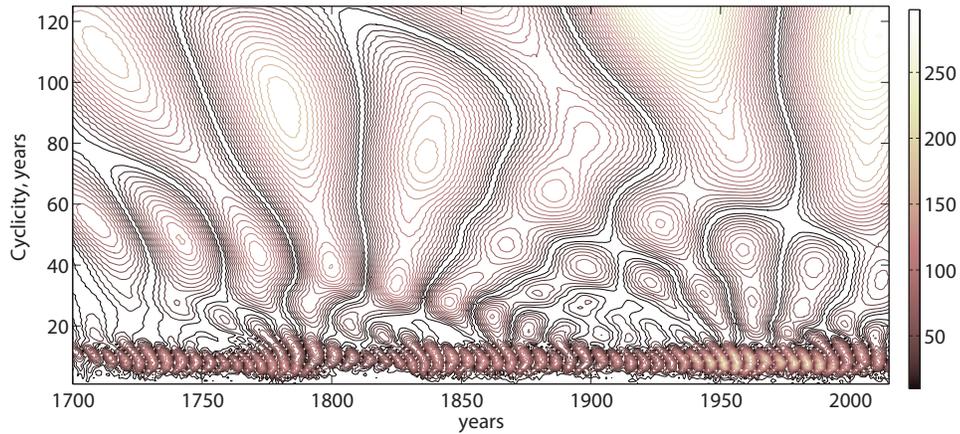}}
 \caption{Cyclic activity of relative sunspot numbers at different time scales.}
{\label{Fi:Fig2}}
\end{figure}

In Fig.2 we can see that periods of cycles on different time scales
are not constant.

The long-term behavior of the sunspot group numbers have been
analyzed using wavelet technique by Frick et al. (1997) who plotted
the changes of the Schwabe cycle (its period is about 11-yr) and
studied the grand minima. The temporal evolution of the Gleissberg
cycle (its period is about 100-yr)can also be seen in the
time-frequency distribution of the solar data. According to Frick et
al. (1997) the Gleissberg cycle is as variable as the Schwabe cycle.
It has two higher amplitude occurrences: first around 1800 (during
the Dalton minimum), and then around 1950. They found very
interesting fact - the continuous decrease in the frequency
(increase of period) of Gleissberg cycle. While near 1750 the cycle
length was about 50 yr, it lengthened to approximately 130 yr by
1950.

In the late of XX century some of solar physicists began to examine
with different methods the variations of relative sunspot numbers
not only in high amplitude 11-yr Schwabe cycle but in low amplitude
cycles approximately equal to half (5.5-yr) and fourth
 (quasi-biennial) parts of period of the main 11-yr cycle, see Vitinsky et al. (1986).
The periods of the quasi-biennial cycles vary considerably within
one 11-yr cycle, decreasing from 3.5 to 2 yrs, and this fact
complicates the study of such periodicity using the method of
periodogram estimates.

Using the methods of frequency analysis of signals  the
quasi-biennial cycles have been  studied not only for the relative
sunspot number, but also for 10.7 cm solar radio emission and for
some other indices of solar activity, see Bruevich et al. (2014b);
Bruevich and Yakunina (2015b). It was also shown that the cyclicity
on the quasi-biennial time scale takes place often among the stars
with 11-yr cyclicity, see Bruevich and Kononovich (2011).

\begin{figure}[tbh!]
\centerline{
\includegraphics[width=140mm]{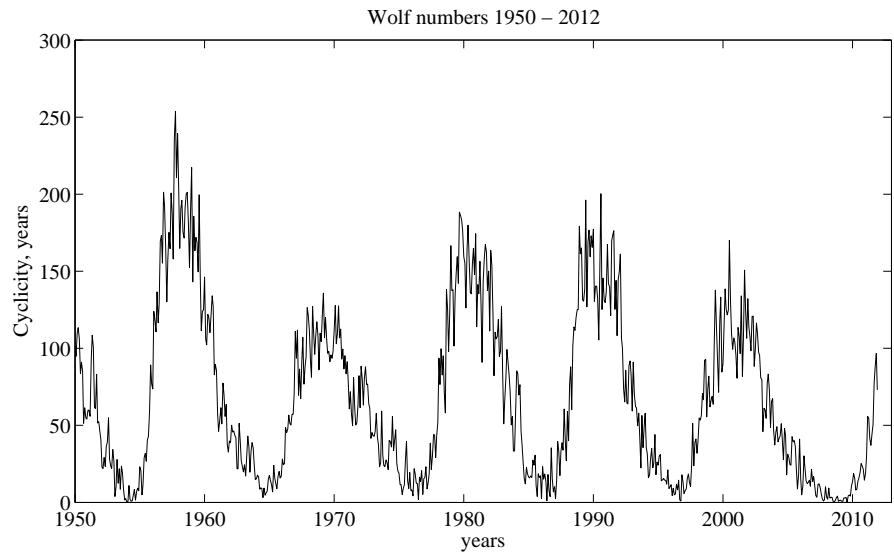}}
 \caption{Monthly averaged relative sunspot numbers 1950 - 2015.}
{\label{Fi:Fig3}}
\end{figure}

For the wavelet analysis of relative sunspot numbers on the scales
in 11 years and quasi-biennial scales we will use the monthly
averaged values, see Fig.3.

\begin{figure}[tbh!]
\centerline{
\includegraphics[width=140mm]{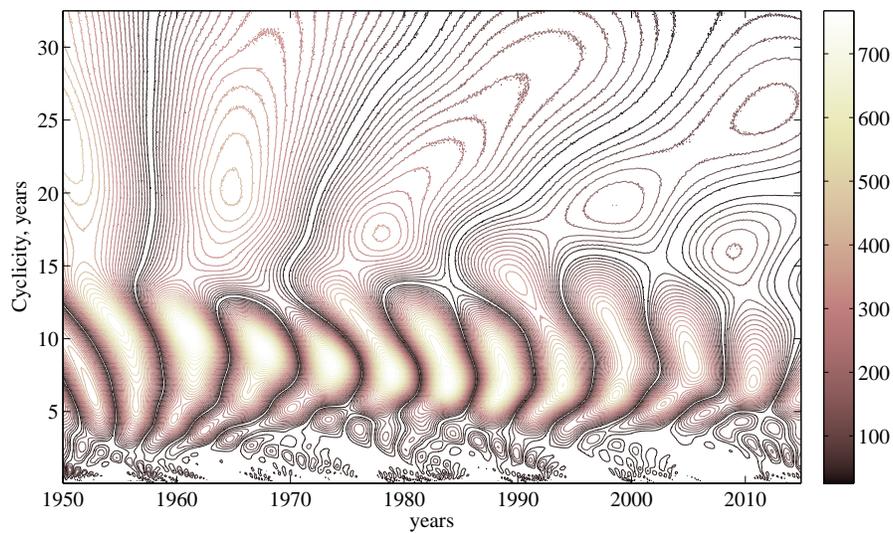}}
 \caption{Cyclic activity of relative sunspot numbers in  activity cycles 18 - 24.}
{\label{Fi:Fig2}}
\end{figure}

In Fig.4 we can see that the periods of cycles on different time
scales are not constant too.

Also as in the case of the learning of the Schwabe cycle we see that
approximately during three cycles value of the periods  decreases
(for the Gleissberg cycle from the periods of 110 years  to 70 years
- Fig.2, for the Schwabe cycle from the periods of 12 years  to 8
years - Fig. 4). Then during the next cycle there are two equal
amplitude cycles (two Gleissberg cycles with periods which change
from 130 to 60 years - Fig.2, the two Schwabe cycles with periods
which change from 13 to 7 years Fig. 4). In the following activity
cycle only the cycles with the greatest periods remain and then  the
value of the periods gradually decreases over the next three cycles.

\begin{figure}[tbh!]
\centerline{
\includegraphics[width=140mm]{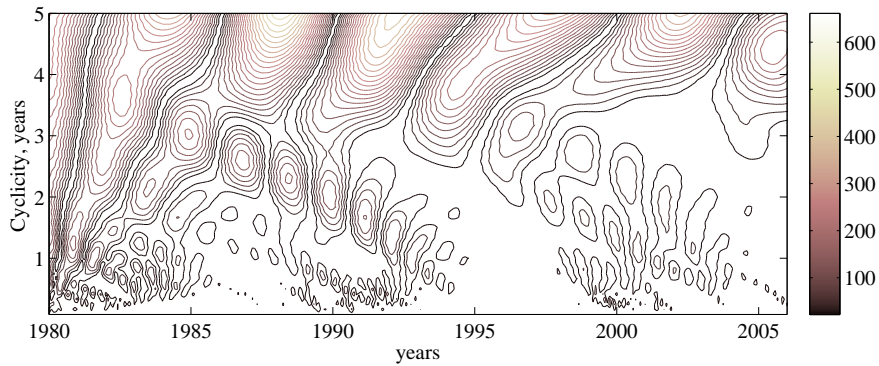}}
 \caption{Cyclic activity of relative sunspot numbers on the quasi biennial time scale.}
{\label{Fi:Fig2}}
\end{figure}

In Fig.5 we can see that in the case of quasi-biennial cycles the
behavior of these periods inside the 11-yr cycle is similar to the
variation of cycle's periods of the Schwabe cycle and the Gleissberg
cycle. The periods of quasi-biennial cycles change from 3.5 to 2 yr
inside the 11-yr cycle.

For the solar-type F,G and K stars according to {\it Kepler}
observations it was also found "shorter" chromosphere cycles with
periods of about two years, see Metcalfe et al. (2010), Garcia et
al. (2010). In  Kollath and Olah (2009) the "shorter" cycle (like
solar quasi-biennial) was determined for the star V CVn, it's
duration is equal to 2.7 yr.

To describe this general trend we propose a formal representation of
this process. The cyclic variations of fluxes of solar radiation (in
particular, the SSN as the most frequently studied activity index)
can be represented by a sinusoid with varying period and constant
amplitude:

$$ A(t) = cos (2\pi\frac{t-t_0}{T}) $$

Note that exactly this behavior we can see in different cycles of
activity, see Fig.2, Fig.4, Fig.5.

The smooth change of the cycle period can be represented as follows:

$$ T(t)=T_0 - k(t) \cdot (t-t_0) $$

where $~t_0$ is the peak time of the studied cycle, $T_0$ is the
cycle's period at the time $t_0$, t varies in the range $t_0<t<
t_0+T_0$.

\begin{table}[tbh!]
\caption{Parameters of different solar cycles.} \vskip12pt

\begin{tabular}{clclcl}

\hline \hline
&&&      \\
~~~~Cyclicity~~~&~~~Cycle's period~~~&~~~ k(t)  \\
\hline
&&&      \\
~~~Century cycle~~~&~~~~~100 yr~~~&~~~~0.3~\\
\hline
&&&      \\
~~~Half a century cycle~&~~~~~ 50 yr~~~&~~~~~0.25~\\
\hline
&&&      \\
~~~11-yr cycle~~~ &~~~~~10~-11 yr~~ &~~~~ 0.2~  \\
\hline
&&&      \\
~~~Quasi biennial cycle~~~&~~~~~ 2 - 3.5 yr~~~&~~~~~0.33~\\
  \hline\hline
\end{tabular}
\end{table}

In Table 1. we presented the values of k(t) for different solar
cycles. For each cycle (from the quasi biennial duration to 11-yr
and 100-yr cycle's periods) the values of coefficient k(t) are
different, see also Fig.2, Fig.3, Fig.4. We consider that it is
necessary to take into account the temporal evolution of solar
cycles for successful forecasts or the parameters of activity
cycles.

\vskip12pt
\section{Conclusion}
\vskip12pt

The study of the evolution of solar cyclicity by example of the SSN
variation using the wavelet analysis allows us to make more accurate
predictions of indices of solar activity (and consequently the
predictions of the parameters of the earth's atmosphere), and also
to take a step towards a greater understanding of the nature of
cyclicity of solar activity. The close interconnection between
activity indices make possible new capabilities in the solar
activity indices forecasts. For a long time the scientists were
interested in the simulation of processes in the earth's ionosphere
and upper atmosphere. For these purposes it is necessary the
successful forecasts of maximum values and other parameters of
future activity cycles and also it has been required to take into
account the century component.

Wavelet analysis of these data reveals the following features: the
period and phase of these relatively low frequency variations of the
solar flux, previous to the studied time point, influence to the
amplitudes and to the phase of studied solar flux which show the
gradually changing of their values in time: as a result, the periods
of variations are getting longer.

\end{document}